\documentclass[aps,prl,superscriptaddress,floatfix,10pt,twocolumn]{revtex4-1}
\usepackage{graphicx,graphics}
\usepackage{dcolumn}
\usepackage{amsmath,amssymb,amsfonts}
\usepackage{latexsym,verbatim}
\usepackage{physics}
\usepackage{bm}
\usepackage{color}
\usepackage{cancel}
\usepackage{ulem}
\usepackage{multirow}
\usepackage[abs]{overpic}
\usepackage[breaklinks=true,colorlinks,citecolor=blue,linkcolor=blue,urlcolor=blue]{hyperref}
\usepackage{overpic}
\usepackage{array}
\usepackage{nicefrac}

\begin{document}
\title{Heat-charge separation in a hybrid superconducting quantum Hall setup}
\author{Carlo Panu}
\affiliation{Dipartimento di Fisica dell'Universit\`a di Pisa, Largo Bruno Pontecorvo 3, I-56127 Pisa,~Italy}
\affiliation{Institute of Condensed Matter Theory and Solid State Optics, Friedrich-Schiller-Universit\"at Jena,
Max-Wien-Platz 1, 07743 Jena,~Germany}
\author{Fabio Taddei}
\affiliation{NEST, Istituto Nanoscienze-CNR and Scuola Normale Superiore, I-56126 Pisa,~Italy}
\author{Marco Polini}
\affiliation{Dipartimento di Fisica dell'Universit\`a di Pisa, Largo Bruno Pontecorvo 3, I-56127 Pisa,~Italy}
\affiliation{ICFO-Institut de Ci\`{e}ncies Fot\`{o}niques, The Barcelona Institute of Science and Technology, Av. Carl Friedrich Gauss 3, 08860 Castelldefels (Barcelona),~Spain}
\author{Amir Yacoby}
\affiliation{Department of Physics, Harvard University, Cambridge, Massachusetts 02138, USA}

\begin{abstract}
Separating heat from charge in a material is an extremely challenging task since they are transported by the very same carriers, i.e.~electrons or holes. In this Letter we show that such separation can reach $100\%$ efficiency in a hybrid superconducting quantum Hall setup, provided that the quantum Hall system is tuned to integer filling factor. We present microscopic calculations for a three-terminal setup to illustrate our idea.
\end{abstract}

\maketitle

{\it {\color{blue}Introduction.---}}Hybrid systems combining the quantum Hall (QH) effect and superconductivity have been the subject of investigation since the turn of the new century~\cite{Ma1993,Takayanagi1998,Takagaki1998,Moore1999,Hoppe2000,Asano2000,Chtchelkatchev2001}. In the absence of a magnetic field, the microscopic mechanism responsible for charge transport at a normal/superconductor interface is the {\it Andreev reflection}, which accounts for the transfer of a Cooper pair into the superconductor (S)~\cite{Tinkham}. Early experimental~\cite{Takayanagi1998,Moore1999} and theoretical~\cite{Takagaki1998,Hoppe2000,Asano2000,Chtchelkatchev2001} research on hybrid QH/S interfaces therefore focused on understanding the peculiarities of charge transport across this interface, possibly stemming from the chiral edge states~\cite{ChangRMP} flowing at the boundaries of a QH fluid. 

More recently, the experimental realization of QH/S hybrid systems has been reported by several groups~\cite{Ermos2005,Batov2007,Komatsu2012,Rickhaus2012,Calado2015,BenShalom2016,Amet2016,Lee2017,Park2017,Draelos2018,Sahu2018,Guiducci2018,Seredinski2019,Zhi2019,Indolese2020,Zhao2020,Wang2021,Hatefipour2022,Gul2022,Barrier2024,Zhao2023,Zhao2023b,Vignaud2023,Hatefipour2023}. 
\begin{figure}[t]
\centering
\includegraphics[width=0.8\columnwidth]{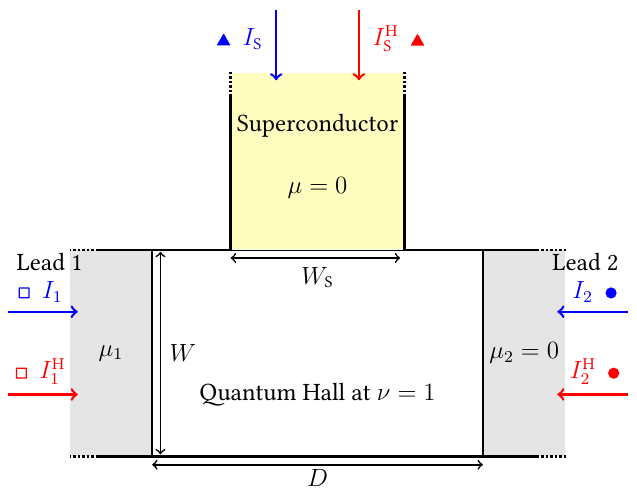}
\caption{(Color online) A three-terminal device where a 2DES in the quantum Hall regime (at filling factor $\nu=1$) is proximitized by an $s$-wave superconductor. A  bias voltage $V$ is applied between leads 1 and 2, with chemical potentials $\mu_1=eV$ and $\mu_2=0$, while the superconducting (S) lead is grounded ($\mu=0$). All leads are assumed to be at the same equilibrium temperature $T$. A quantizing magnetic field $B\neq 0$ is present in the white and grey regions (with filling factor $\nu=1$, spin polarized), while in the superconducting lead (colored in yellow) the field is set to zero. Charge ($I_{1,2,{\rm S}}$) and heat ($I^{\text{H}}_{1,2,{\rm S}}$) currents are calculated as entering the central region. Charge currents will be denoted by blue symbols throughout this Letter: empty squares for lead 1, filled circles for lead 2, and filled triangles for the superconducting lead. Heat currents will instead be denoted by red symbols (as in the case of charge currents, empty squares for lead 1, filled circles for lead 2, and filled triangles for the superconducting lead).
\label{setup}}
\end{figure}
The vast majority of experimental realizations is based on graphene QH systems~\cite{Komatsu2012,Rickhaus2012,Calado2015,BenShalom2016,Amet2016,Lee2017,Park2017,Draelos2018,Sahu2018,Seredinski2019,Indolese2020,Wang2021,Zhao2020,Gul2022,Barrier2024,Vignaud2023,Zhao2023,Zhao2023b}.
Indeed, graphene encapsulated in hexagonal boron nitride allows the fabrication of high-quality contacts~\cite{epitaxy} with high-critical-field superconductors such as MoRe, NbN, MoGe, NbSe$_2$.

Our aim in this Letter is to achieve separation of charge and heat flows in an experimentally relevant solid-state platform. This is very well known to be a truly challenging task since heat-charge separation~\cite{nota-hcsep,nota-Kh} requires the violation of the Wiedemann-Franz law~\cite{Aschcroft1976}. Such violations are rare in nature, and typically require e.g. strong electron-electron interactions (see e.g.~Refs.~\cite{Wakeham2011,Principi2015,Buccheri2022,Crossno2016,Buccheri2022} and references therein).
We achieve this challenging goal by using three key ingredients:
1) The first key idea is to bypass the Wiedemann-Franz law by using a QH/S setup where the QH system is at filling factor $\nu=1$. The fact that superconductors are poor heat conductors will play a crucial role below. These hybrid QH/S setups are currently at the center of a great deal of attention because in certain regimes are expected to support novel non-Abelian excitations~\cite{Mong2014,Clarke2014}, relevant for topological quantum computation~\cite{Nayak2008,Lahtinen2017}. Being able to manage heat~\cite{Giazotto2006,Muhonen2012,Cahill2014,Benenti2017} in such devices, where quantum computation relies on fragile quasiparticles, is of critical importance. 
2) The second key element that is required to separate heat from charge is to have a device with three terminals, as is the case in the one we propose, which is sketched in Fig.~\ref{setup}. 
3) The third key element we use is {\it spin mixing}. Indeed, the co-existence of QH physics and superconductivity requires high-critical-field superconductors, which tend to be endowed with strong spin-orbit coupling and disorder. These two elements are crucial as they induce spin-mixing at the QH/S interface, which would be otherwise negligible. Without spin-mixing, indeed, the superconducting terminal of our $\nu=1$ QH/S three-terminal device would be totally {\it inert}, in the sense that charge and heat currents would be perfectly coupled, flowing entirely between the other two (normal) leads. Here we demonstrate that spin-mixing in the superconducting terminal leads to a perfect balance between normal and Andreev transmission in a {\it finite range} of energies.

The actual setup we consider (Fig.~\ref{setup}) is an inverse-T shaped two-dimensional electron system (2DES) subject to a quantizing uniform and perpendicular magnetic field $B$. The horizontal section of the device has a width $W$, while the vertical section has a width $W_{\rm S}$. Two electrodes (labelled 1 and 2, in grey) are attached to the sides and a superconductor (in yellow) is deposited on the top section of the 2DES, so that the latter acquires ``superconducting properties'', i.e.~a finite order parameter $\Delta$ through the proximity effect. In addition to $W$ and $W_{\rm S}$, our device is characterized by a third length scale associated with $\Delta$, i.e.~the superconducting coherence length $\xi=\hbar v_{\rm F}/(\pi\Delta)$, where $v_{\rm F}$ is the 2D, bulk Fermi velocity. We assume that the magnetic field is completely expelled from the proximitized region due to the Meissner effect~\cite{Caveat_Meissner_Effect}. A  bias voltage $V$ is applied between leads 1 and 2, with chemical potentials $\mu_1=eV$ and $\mu_2=0$, while the superconducting lead is grounded (the superconducting condensate chemical potential $\mu=0$).

{\color{blue}{\it Theory of heat-charge separation in a hybrid QH/S system.---}}
The system is characterized by five energy scales: the cyclotron gap $\hbar\omega_{\rm c}$, the thermal energy $k_{\rm B} T$, the Zeeman splitting $g\mu_{\rm B} B$, the superconducting gap $\Delta$, and the bias voltage $eV$. Here, $g$ is the Land\'e factor, $\mu_{\rm B}$ the Bohr magneton, and $B$ the intensity of the applied perpendicular magnetic field.
Given a certain electron density in the 2DES, we set a value of the magnetic field $B$ in order for transport to be mediated by a single, spin-polarized, edge state. This occurs when the 2DES is tuned at filling factor $\nu=1$, i.e.~when the chemical potential of the 2DES sits between the first and the second Landau level.
Moreover, we choose the following working conditions:
\begin{enumerate}
\item[i)]  the chemical potential of the leads are $\mu_1=eV < \Delta$ and $\mu_2=\mu=0$, while temperature is such that $k_{\rm B} T \ll \Delta$, i.e.~we are in the sub-gap transport regime;
\item[ii)]  $eV <  \, g \mu_{\rm B} B$, i.e~the chemical potential difference between leads $1$ and $2$ is smaller than the Zeeman splitting energy;

\item[iii)]  $k_{\rm B} T \ll g \mu_{\rm B} B$, i.e.~the thermal energy is smaller the Zeeman splitting energy.
\end{enumerate}
Notice that conditions i) to iii) ensure that a single spin-polarized edge channel is available for transport. Because of that, no Andreev processes can occur since a Cooper pair in an $s$-wave superconductor is made of electrons of {\it both} spin species.
Andreev-mediated transport at $\nu=1$, however, has been observed in recent experiments~\cite{Lee2017,Gul2022}, where NbN was employed as a superconductor, and attributed to spin-flipping scattering mechanisms.
Indeed, in Ref.~\onlinecite{Wakamura2014} it was shown that such processes can take place in NbN because of the presence of strong spin-orbit coupling.

Many authors have addressed theoretically transport in hybrid QH/S systems~\cite{Giazotto2005,Khaymovich2010,Stone2011,Ostaay2011,Gamayun2017,Beconcini2018,Finocchiaro2018,Sekera2018,Zhang2019,Huang2019,Gavensky2020,Galambos2022,Manesco2022,Clem2010,Tang2022,Kurilovich2022,Khrapai2023,Kurilovich2023,David2023,Schiller2023,Michelsen2023,Blasi2023,Cuozzo2023,Arrachea2023}.
Here, we assume coherent transport and calculate charge and heat currents flowing in the electrodes within the Landauer-B\"uttiker scattering approach~\cite{Lambert1998}.
Because of the chiral nature of edge states,
they can be expressed only through two sets of transmission coefficients: a) the {\it  normal} transmission coefficients ${\cal T}_{\alpha,\alpha}(E)$ and ${\cal T}'_{\alpha,\alpha}(E)$ with ${\cal T}_{\alpha,\alpha}(E)$ [${\cal T}'_{\alpha,\alpha}(E)$] being the probability for an $\alpha$-type particle ($\alpha=+$ for electrons and $\alpha=-$ for holes) at energy $E$ to be transferred from lead 1 to  2 [from lead 2 to 1]; b) the {\it Andreev} transmission coefficients ${\cal T}_{\alpha,\beta}(E)$ and ${\cal T}'_{\alpha,\beta}(E)$, with $\alpha\ne\beta$, where ${\cal T}_{\alpha,\beta}(E)$ [${\cal T}'_{\alpha,\beta}(E)$] representing the probability for a particle of type $\beta$ at energy $E$ which starts from lead $1$ [$2$] to be converted into a particle of type $\alpha\neq \beta$ when reaching lead $2$ [$1$].
We recall that Andreev scattering processes are the ones responsible for the charge transfer at a normal/S interface~\cite{Tinkham,Blonder1982}.

The charge currents in the normal leads, assuming sub-gap transport regime, can be written as
\begin{equation}
\label{chcurr1}
I_{1} = \frac{e}{h}\int^{\infty}_{0}dE \Bigl\{ \sum_{\alpha,\beta}(\alpha) {\cal T}'_{\alpha,\beta}(E) \bigl[f_1^{\alpha}(E)-f_2^{\beta}(E)\bigr]
 \Bigr\}~,
\end{equation}
and
\begin{equation}
\label{chcurr2}
I_{2} = \frac{e}{h}\int^{\infty}_{0}dE \Bigl\{ \sum_{\alpha,\beta}(\alpha) {\cal T}_{\alpha,\beta}(E) \bigl[f_2^{\alpha}(E) - f_1^{\beta}(E)\bigr]  \Bigr\}~.
\end{equation}
Here, 
\begin{equation}
f^{\pm}_i(E)=\frac{1}{{\rm exp}\big\{\big[E \mp (\mu_i-\mu)\big]/k_{\rm B} T\big\}+1}
\end{equation}
is the Fermi-Dirac distribution function for electrons/holes at energy $E$ in lead $i=1,2$, evaluated at the equilibrium temperature $T$. 
Notice that in writing Eqs.~(\ref{chcurr1})-(\ref{chcurr2}) we have used the fact that the reflection probabilities at the two leads vanish, because of the chiral character of quantum Hall edge states. Under the same assumption, the heat currents can be calculated from the relation~\cite{Benenti2017}:
\begin{equation}
\label{eq:defIH}
    I^{\rm H}_i = I^{\rm E}_i- \frac{\mu_i}{e} I_i~,  
\end{equation}
where $I^{\rm E}_i$ is the energy current in the $i$-th lead:
\begin{equation}
\label{encurr10}
I^{\rm E}_{1} = \frac{1}{h}\int^{\infty}_{0}dE\, E \Bigl\{ \sum_{\alpha,\beta} {\cal T}'_{\alpha,\beta}(E) \bigl[f_1^{\alpha}(E) - f_2^{\beta}(E)\bigr]  \Bigr\}~,
\end{equation}
and
\begin{equation}
\label{encurr2}
I^{\rm E}_{2} = \frac{1}{h}\int^{\infty}_{0}dE\, E \Bigl\{ \sum_{\alpha,\beta} {\cal T}_{\alpha,\beta}(E) \bigl[f_2^{\alpha}(E) - f_1^{\beta}(E)\bigr]  \Bigr\}~.
\end{equation}
Two comments are now in order: a) The second term in Eq.~(\ref{eq:defIH}), which is proportional to the chemical potential $\mu_i$ in the $i$-th lead, stems from the fact that an electron with energy $E$ leaving the $i$-th reservoir carries away an amount of heat $\Delta Q_i = E- \mu_i $. b) It should be noted here that one needs $W_{\rm S}\gtrsim\xi$ in order for the Andreev transmission coefficients to be finite.
In fact, the superconducting coherence length $\xi$ represents the minimum length necessary for superconducting correlations to develop.

In order to maximally simplify the next formulas, we define the Andreev transmission coefficients ${\cal T}_{\rm A}(E) \equiv {\cal T}_{-,+}(E)$ and ${\cal T}'_{\rm A}(E)\equiv {\cal T}'_{-,+}(E)$ and the normal transmission coefficients ${\cal T}_{\rm N}(E)\equiv {\cal T}_{+,+}(E)$ and ${\cal T}'_{\rm N}(E) \equiv {\cal T}'_{+,+}(E)$. Due to the chiral nature of edge states, in the energy range where $f^{\pm}_i(E)$ are finite one expects ${\cal T}^{\prime}_{\rm N}(E)=1$ and ${\cal T}^{\prime}_{\rm A}(E)=0$, because electrons that start from lead 2 can only propagate (as electrons) on the lower edge, where there is no superconducting proximity effect, with no possibility of Andreev scattering---see Fig.~\ref{setup}.
Therefore, in the zero-temperature limit, and under the above operating conditions, the expressions for the charge and heat currents greatly simplify, reducing to:
\begin{equation}
\label{chcurr1T}
    I_{1}  = \frac{e^2}{h} V ~,
\end{equation}
\begin{equation}
\label{chcurr2T}
    I_{2} = \frac{e}{h}\int^{eV}_{0}dE \Bigl[{\cal T}_{\rm A}(E)-{\cal T}_{\rm N}(E)\Bigr]~,
\end{equation}
\begin{equation}
\label{heatcurr1T}
I^{\rm{H}}_{1} = -\frac{(eV)^2}{2 h} ~,
\end{equation}
and
\begin{equation}
\label{heatcurr2T}
I^{\rm{H}}_{2} = -\frac{1}{h} \int^{eV}_{0}dE~E\Bigl[{\cal T}_{\rm N}(E)+{\cal T}_{\rm A}(E)\Bigr]~.
\end{equation}
Heat and charge currents flowing in the superconducting lead can be determined from the conservation of particle and energy currents, namely
\begin{equation}
    I_1 + I_2 + I_{\rm S} = 0~,
\end{equation}
and
\begin{equation}
\label{eq:consE}
    I^{\rm E}_1 + I^{\rm E}_2 + I^{\rm E}_{\rm S} = 0~.
\end{equation}
(Notice that currents entering in the central region, i.e.~in the white area in Fig.~\ref{setup}, are defined to be positive.)
Using Eqs.~(\ref{eq:defIH})  and (\ref{eq:consE}) we get
\begin{equation}
\label{heatcurrSCT}
I^{\rm H}_{\rm S} = -I^{\rm H}_1 - I^{\rm H}_2 - V I_1~.
\end{equation}

We now hypothesize that a region of parameter space exists where normal and Andreev transmissions balance, in the relevant range of energies, i.e.~that a region of parameters exists such that ${\cal T}_{\rm A}(E)={\cal T}_{\rm N}(E)=1/2$ in the energy range $E\in [0, eV]$.
Replacing this equality in Eq.~(\ref{chcurr2T}) one finds $I_2 = 0$. In the same limit, combining Eq.~(\ref{heatcurr2T}) with Eqs.~(\ref{chcurr1T}), (\ref{heatcurr1T}), and (\ref{heatcurrSCT}), we get $I_{\rm S}^ {\rm H}= 0$. This means that in lead $2$ the only non-zero current is the heat current, while in the superconducting lead the only non-zero current is the charge current.
This realizes the {\it separation of heat and charge}.
Indeed, the charge and heat currents flowing in lead 1 (both finite) are spatially separated in the sense that heat current is diverted into lead $2$ while charge current is diverted into the S lead. This is the most important prediction of this Letter and it applies to hybrid QH/S devices, {\it independently} of the microscopic band structure of the 2DES. The only requirements are that the 2DES is in the QH regime at $\nu=1$ and that the operating conditions i)-iii) above are satisfied.
For completeness, one also finds  $I^{\rm H}_2= I^{\rm H}_1 = -(eV)^2/(2 h) $, while $I_{\rm S} \simeq -e^2V/h$.
We notice that the relation $I_{\rm S}^ {\rm H}= 0$ is not surprising since superconductors are poor heat conductors for temperatures below the gap.
In superconducting junctions, heat flow is intrinsically suppressed for low enough voltages and temperatures.

One may argue that achieving perfect balance between normal and Andreev transmission in the whole range of energies $E\in [0, eV]$ is implausible as it would require tremendous fine tuning. According to Ref.~\onlinecite{Kurilovich2022}, however, ${\cal T}_{\rm A}(E)={\cal T}_{\rm N}(E)=1/2$ actually occurs in the case of two counter-propagating $\nu = 1$ QH edge states, provided that they are coupled through a narrow ``dirty'' superconducting electrode in which {\it spin flipping} is allowed.
More precisely, in this configuration, the amplitudes of crossed Andreev reflection and elastic cotunelling processes through the narrow superconductor are random and are statistically balanced. In what follows, we prove with a numerical calculation that the relation ${\cal T}_{\rm A}(E)={\cal T}_{\rm N}(E)=1/2$ for $E\in [0, eV]$ also holds for our setup.
\begin{figure}[t]
	\centering
	\includegraphics[width=0.9\columnwidth]{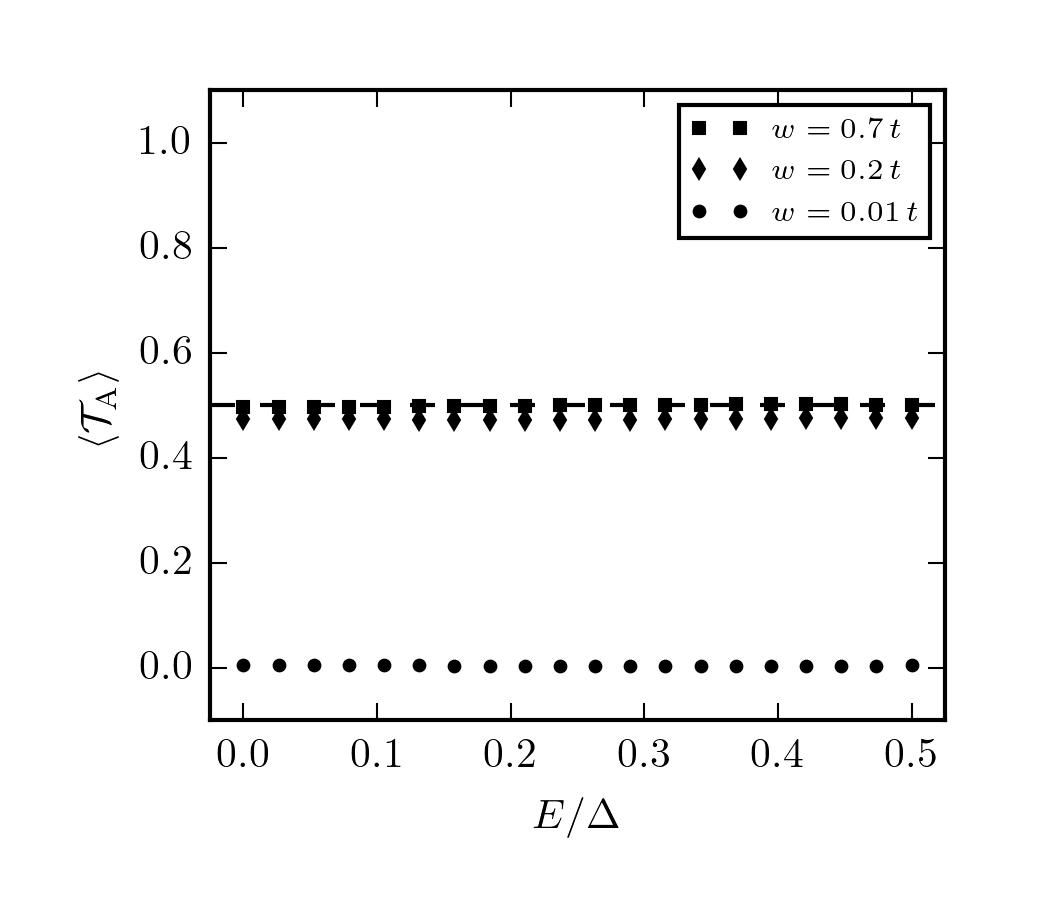}
	\caption{The ensemble-averaged Andreev transmission $\langle{\cal T}_{\rm A}\rangle$ between lead 1 and 2, for the conversion of electrons into holes, is plotted as a function of energy $E$ (measured in units of $\Delta$). A dashed horizontal line indicates the value $\langle{\cal T}_{\rm A}\rangle=1/2$.  Uniformly distributed spin-mixing disorder is present only in the superconducting region. Data in this figure correspond to three values of the disorder strength, $w= 0.01t$ (circles), $w =0.2t$ (diamonds), and $w =0.7t$ (squares), and have been obtained by averaging over $2000$ disorder configurations. Other parameters used in the calculations are: $\Delta =1.5~{\rm meV}$, $\nu=1$, $B=5~{\rm T}$, $W = 247~{\rm nm}$, $W_{\rm S} = 330~{\rm nm}$, $t= 0.1~{\rm eV}$, $\varepsilon=0.392~{\rm eV}$, $m = 0.035~m_{\rm e}$ (where $m_{\rm e}$ is the bare electron mass in vacuum), and Land\'e factor $g=20$. With these parameters, we obtain $\hbar \omega_{\rm c} \simeq 16~{\rm meV}$, $g\mu_{\rm B} B \simeq 6~{\rm meV}$, and superconducting coherence length $\xi\simeq 40~{\rm nm}$. All the parameters and corresponding symbols in this caption have been defined in Ref.~\cite{SM}.}
	\label{TA}
\end{figure}

{\it {\color{blue}Numerical example.---}}In order to calculate numerically the transmission probabilities ${\cal T}_{\rm A}$ and ${\cal T}_{\rm N}$ as functions of energy, we model the hybrid system in Fig.~\ref{setup} with a discretized Bogoliubov-De Gennes Hamiltonian~\cite{deGennes}, which is explicitly described in Ref.~\cite{SM}. The transmission probabilities are calculated numerically, using the  KWANT~\cite{kwant} toolkit, carrying out averages over a large number of disorder realizations.

Fig.~\ref{TA} shows the ensemble-averaged Andreev transmission probability $\langle {\cal T}_{\rm A}\rangle$ as a function of the energy $E$ in units of the superconducting gap $\Delta$, for three different values of the disorder strength $w$ (the horizontal dashed line indicates the value $1/2$).
We have used values of the parameters which satisfy the conditions i) to iii) listed earlier~\cite{nota-iiii}.
In agreement with Ref.~\cite{Kurilovich2022}, $\langle {\cal T}_{\rm A}(E)\rangle$ is practically pinned at the value $\langle {\cal T}_{\rm A}(E)\rangle=1/2$ in the whole energy range of the plot as long as disorder is strong enough, i.e.~for $w \gtrsim 0.7t$ (squares).
In the case of a weaker disorder, i.e.~for $w=0.2t$ (diamonds), the value of $\langle {\cal T}_{\rm A}(E)\rangle$ is just slightly below $1/2$, while for very weak disorder, i.e.~for~$w =0.01t$ (circles), we get, as expected, $\langle {\cal T}_{\rm A}(E)\rangle\simeq 0$~\cite{footGlazman}.
\begin{figure}[t]
	\centering
	\includegraphics[width=0.9\columnwidth]{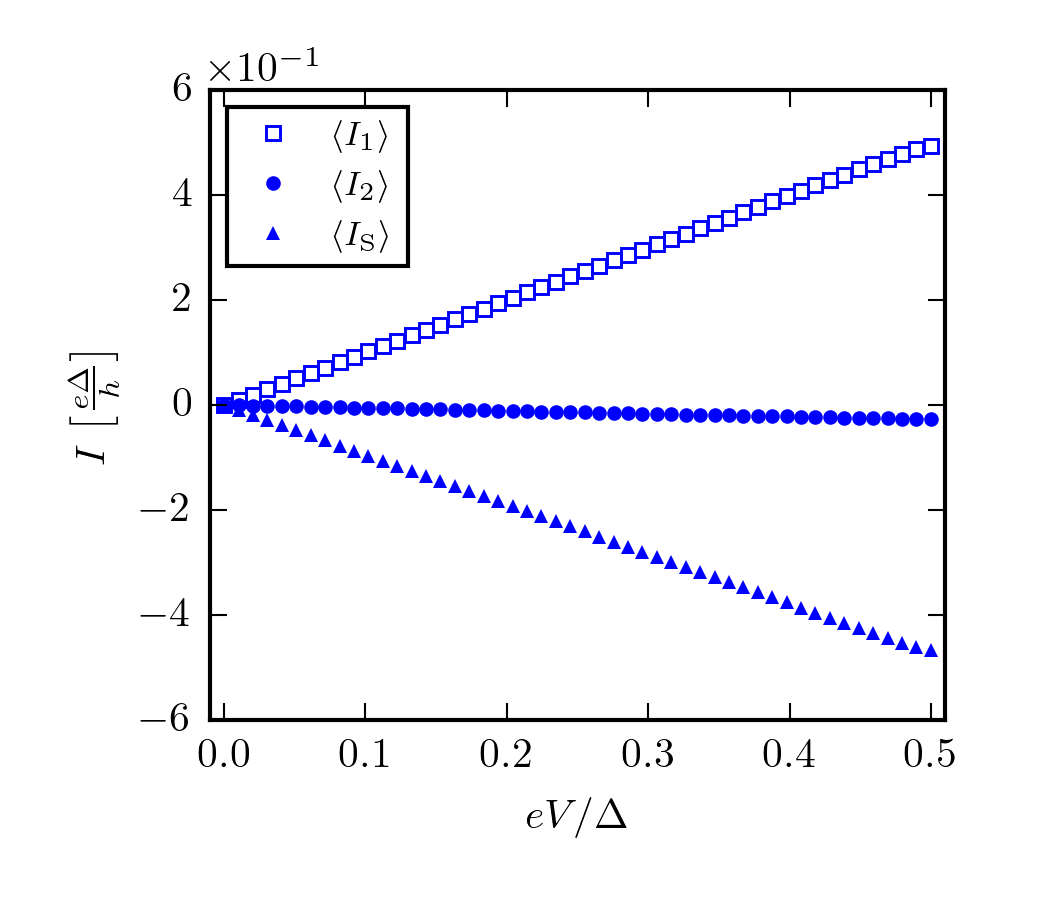}
	\caption{(Color online) Ensemble-averaged charge currents (in units of $e\Delta/h$) flowing in lead $1$ (squares), lead $2$ (circles), and in the superconducting lead (triangles) are plotted as functions of voltage $V$ (in units of $\Delta/e$). Results in this plot refer to a temperature $T=0.01~\Delta/k_{\rm B}$ and $w =0.2t$. Other parameters are identical to those reported in Fig.~\ref{TA}. Note that the charge current $\langle I_2\rangle$ in lead $2$ is vanishingly small: the current is taken away by the superconductor.
	 \label{Ich}}
\end{figure}

\begin{figure}[t]
\centering
\includegraphics[width=0.9\columnwidth]{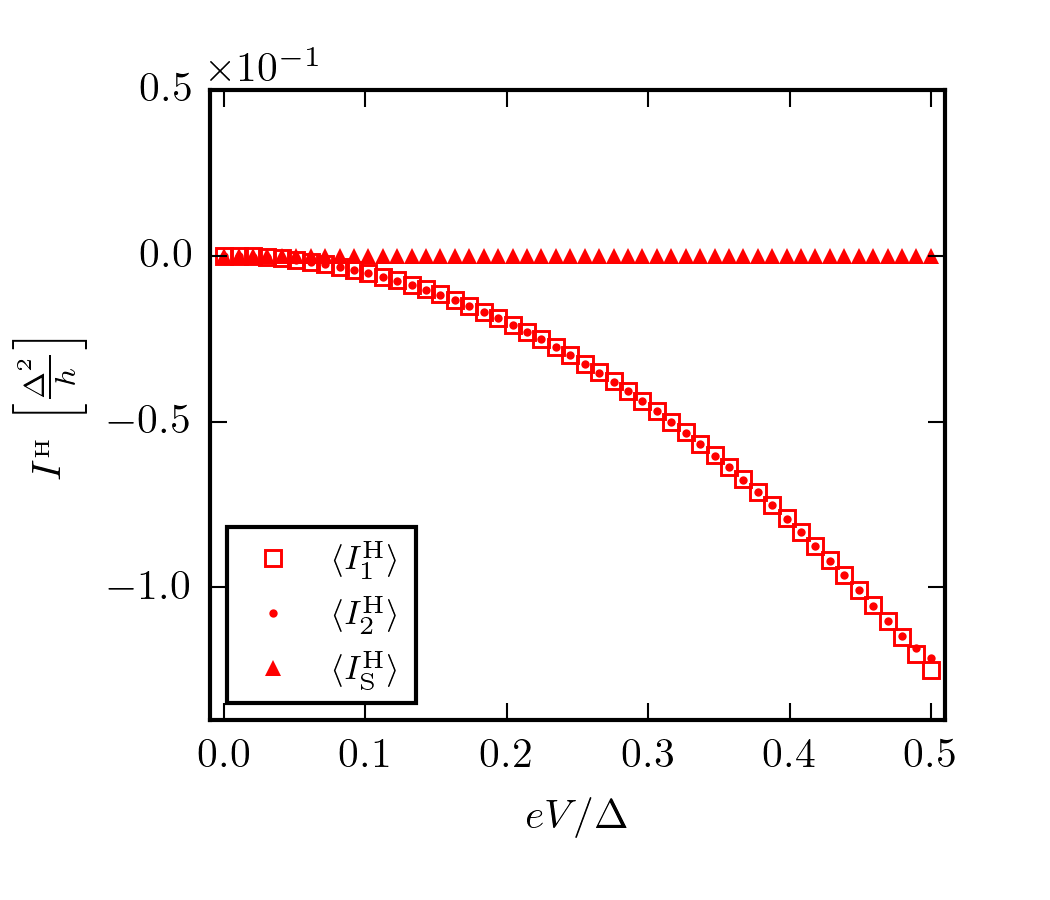}
\caption{(Color online) Ensemble-averaged heat currents flowing in lead 1 (squares), lead 2 (circles), and in the superconducting lead (triangles) are plotted as functions of voltage $V$ (in units of $\Delta/e$). The parameters used for obtaining these results coincide with those reported in the captions of Figs.~\ref{TA} and~\ref{Ich}. Heat currents are finite in leads 1 and 2, but vanish in the superconducting lead.
\label{Iq}}
\end{figure}

The resulting charge and heat currents, for the case $w=0.2t$, are plotted in Figs.~\ref{Ich} and \ref{Iq}, respectively, as functions of the voltage $V$ and for a very low temperature, i.e.~$T=1.0\times 10^{-2}\Delta/k_{\rm B}$~\cite{nota-temp}.
In perfect agreement with our theoretical analysis above, Fig.~\ref{Ich} shows that the charge current in lead $2$ (circles) vanishes in the whole range of explored voltages. The charge current flowing in lead 1 (squares) is entirely collected by the superconducting lead (triangles).
It is worth noticing that the currents $\langle I_1\rangle$ and $\langle I_{\rm S}\rangle$ linearly depend on $V$.
As far as the heat current is concerned, Fig.~\ref{Iq} shows that this is zero in the superconducting lead (triangles) while it is finite and negative in both leads 1 (squares) and 2 (circles). The negative sign implies that both heat currents flow away from the central region of the device, thus representing Joule heating.
Moreover, according to Eqs.~(\ref{heatcurr1T}) and (\ref{heatcurr2T}), the two heat currents, $\langle I^{\rm H}_1\rangle$ and $\langle I^{\rm H}_2\rangle$, are approximately equal because $\langle{\cal T}_{\rm N}(E)\rangle\simeq \langle{\cal T}_{\rm A}(E)\rangle\simeq 1/2$.

{\it {\color{blue}Discussion.---}}Results in Figs.~\ref{Ich} and~\ref{Iq} demonstrate that our setup is indeed a three-terminal heat-charge separator. This functionality is a consequence of three crucial ingredients: a) the chiral and spin-polarized nature of $\nu=1$ QH edge states, b) the specific working conditions we have chosen, i.e.~the hierarchy i)-iii) between the various relevant energy scales of the problem, and c) spin mixing in the superconducting region. The latter is expected to be a rather general feature of all superconducting materials with strong spin-orbit coupling~\cite{Wakamura2014,Lee2017,Gul2022}. Careful readers will have noted that our numerical calculations have been carried out for a 2DES with a single parabolic band. They can of course be extended to the case of a graphene QH/S hybrid device as e.g. in Ref.~\cite{Beconcini2018}. The point, however, is that in this case, because of the spin and valley degeneracies in zero magnetic field, a $\nu=1$ QH state is realized thanks to many-body effects~\cite{Goerbig_RMP_2011} (i.e.~exchange effects stemming from long-range electron-electron interactions). These can be taken into account in our Bogoliubov-de Gennes Landauer-B\"{u}ttiker approach by treating electron-electron interactions at the level of the Hartree-Fock approximation~\cite{Novelli_PRL_2019}. While this is certainly interesting, it is well beyond the scope of the present work. The theoretical analysis discussed in the first part of this Letter shows indeed that the phenomenon of heat-charge separation in hybrid  QH/S systems is universal, provided that the three above mentioned crucial ingredients are taken into account. 

{\color{blue}{\it Acknowledgements.---}}This work was supported by the European Union's Horizon 2020 research and innovation programme under the Marie Sklodowska-Curie grant agreement No.~873028 - Hydrotronics, by the MUR - Italian Minister of University and Research under the ``Research projects of relevant national interest  - PRIN 2020''  - Project No.~2020JLZ52N, title ``Light-matter interactions and the collective behavior of quantum 2D materials (q-LIMA)'', by the MUR-PRIN 2022 - Grant No. 2022B9P8LN - (PE3) - Project NEThEQS  
``Non-equilibrium coherent thermal effects in quantum systems'' in PNRR Mission 4 - Component 2 - Investment 1.1 ``Fondo per il Programma Nazionale di Ricerca e Progetti di Rilevante Interesse Nazionale'' (PRIN) funded by the European Union - Next Generation EU, and by the Royal Society through the International Exchanges Scheme between the UK and Italy (Grant No.~IEC R2 192166). 
C. Panu is supported by the International Research Training Group (IRTG) 2675 ``Meta-ACTIVE'', Project No. 437527638.
A. Yacoby is supported by the Quantum Science Center
(QSC), a National Quantum Information Science Research Center of the U.S. Department of Energy (DOE).
A. Yacoby is also partly supported by the Gordon and
Betty Moore Foundation through Grant GBMF 9468, and by
the U.S. Army Research Office (ARO) MURI project under grant number W911NF-21-2-0147.


%

\clearpage 
\newpage

\setcounter{section}{0}
\setcounter{equation}{0}%
\setcounter{figure}{0}%
\setcounter{table}{0}%

\setcounter{page}{1}

\renewcommand{\thetable}{S\arabic{table}}
\renewcommand{\theequation}{S\arabic{equation}}
\renewcommand{\thefigure}{S\arabic{figure}}
\renewcommand{\bibnumfmt}[1]{[S#1]}
\renewcommand{\citenumfont}[1]{S#1}

\onecolumngrid

\begin{center}
\textbf{\Large Supplemental Material for:\\ ``Heat-charge separation in a hybrid superconducting quantum Hall setup''}
\bigskip

Carlo Panu,$^{1,\,2}$
Fabio Taddei,$^{3}$
Marco Polini,$^{1,\,4}$
Amir Yacoby$^{5}$

\bigskip

$^1$\!{\it Dipartimento di Fisica dell'Universit\`a di Pisa, Largo Bruno Pontecorvo 3, I-56127 Pisa,~Italy}

$^2$\!{\it Institute of Condensed Matter Theory and Solid State Optics, Friedrich-Schiller-Universit\"at Jena,
Max-Wien-Platz 1, 07743 Jena,~Germany}

$^3$\!{\it NEST, Istituto Nanoscienze-CNR and Scuola Normale Superiore, I-56126 Pisa,~Italy}

$^4$\!{\it ICFO-Institut de Ci\`{e}ncies Fot\`{o}niques, The Barcelona Institute of Science and Technology, Av. Carl Friedrich Gauss 3, 08860 Castelldefels (Barcelona),~Spain}

$^5$\!{\it Department of Physics, Harvard University, Cambridge, Massachusetts 02138, USA}

\bigskip

In this Supplemental Material we present all the necessary technical details on the Hamiltonian we have used in our numerical calculations.
 
\end{center}

\onecolumngrid

\appendix

The Hamiltonian $\hat{\cal H}_0$ describing the 2DES is discretized on a square lattice, with lattice constant $a$, and given by
\begin{equation}
\hat{\cal H}_{0} = \sum_{i,\sigma}  (\varepsilon+\sigma g\mu_{\rm B} B_i)\, \hat{c}^\dagger_{i\sigma}\hat{c}_{i\sigma} +
\sum_{\langle i,j \rangle,\sigma} t \,e^{i\phi_{ij}}\, \hat{c}^\dagger_{i\sigma}\hat{c}_{j\sigma} + {\rm H.c}~,
\end{equation}
where $\hat{c}^\dagger_{i\sigma}$ is the creation operator for an electron of spin $\sigma$ on site $i$, $\varepsilon$ is the on-site energy, and $t$ is the  tight-binding hopping energy. 
This implies that, at low energies, the 2DES has a single parabolic band with an effective mass given by  $m = \hbar^2/(2t a^2)$.
Moreover, the Zeeman energy is expressed in terms of the Land\'e $g$-factor, the Bohr magneton $\mu_{\rm B}$, and the intensity of the applied uniform perpendicular magnetic field $B_i = B$, non-zero in the white- and grey-colored regions of Fig.~1.
The orbital effect of the latter is accounted for by the Peierls substitution through the complex phase $\phi_{ij}=2\pi\phi_0^{-1}\int_i^j{\bm A}\cdot {\rm d}{\bm l}$, where ${\bm A}=(-By,0,0)$ is the vector potential in the Landau gauge and $\phi_0=h/e$ is the flux quantum (see, for example, Ref.~\cite{Beconcini2018}).
The sum over $\langle i,j \rangle$ runs over neighboring sites.

Superconducting pairing is introduced through the term
\begin{equation}
\hat{\cal H}_{\rm S} = \sum_i  \Delta_i \hat{c}^\dagger_{i\uparrow}\hat{c}^\dagger_{i\downarrow} + {\rm H.c},
\end{equation}
where $\Delta_i$ (the site-dependent superconducting order parameter) is non-zero (and equal to $\Delta$ for all sites) only in the superconducting region, colored in yellow in Fig.~1.
Moreover, in the superconducting region we assume that $B_i$ is equal to zero, i.e.~completely expelled by the Meissner effect.

To account for a uniformly distributed spin-mixing disorder in the superconducting region we add to the Hamiltonian the term 
\begin{equation}
\hat{\cal H}_{\rm SM} = w\sum_i  F_i \hat{c}^\dagger_{i\uparrow}\hat{c}_{i\downarrow} + {\rm H.c},
\end{equation}
where $w$ is the disorder strength and $F_i$ are random numbers, uniformly distributed in the range $[-1,1]$. We set $F_i=0$ on all sites $i$ which do not belong to the superconducting region.

The overall Hamiltonian is therefore $\hat{\cal H}=\hat{\cal H}_{0} +\hat{\cal H}_{\rm S} +\hat{\cal H}_{\rm SM}$, which can be written in the Bogoliubov-De Gennes block matrix form following Refs.~\cite{Ludwig2016, beenakker2011random}.


\begin{thebibliography}{100}
\bibitem{nota-temp}
Note that, because of the presence of a superconductor, experiments have to be conducted at low (cryogenic) temperatures, where phonons are frozen and do not play any role in electronic transport.
\end{thebibliography}
\end{document}